\documentclass[aps,preprintnumbers,pacs,nofootinbib]{revtex4}
\usepackage{graphicx}
\usepackage{amsmath}
\usepackage{amssymb}

\begin{document}

\title{Weyl fluid dark matter model tested on the galactic scale \\
by weak gravitational lensing}
\author{K. C. Wong$^{1}$}
\email{fankywong@gmail.com}
\author{T. Harko$^{2}$}
\email{harko@hkucc.hku.hk}
\author{K. S. Cheng$^{2}$}
\email{hrspksc@hkucc.hku.hk}
\author{L. \'{A}. Gergely$^{3}$}
\email{gergely@physx.u-szeged.hu}
\affiliation{$^{1}$Department of Physics, University of Hong Kong, Pok Fu Lam Road, Hong
Kong, P. R. China\\
$^{2}$Department of Physics and Center for Theoretical and Computational
Physics, The University of Hong Kong, Pok Fu Lam Road, Hong Kong, P. R. China%
\\
$^{3}$Departments of Theoretical and Experimental Physics, University of
Szeged, D\'{o}m T\'{e}r 9, Szeged 6720, Hungary}

\begin{abstract}
The higher dimensional Weyl curvature induces on the brane a new source of
gravity. This Weyl fluid of geometrical origin (reducing in the spherically
symmetric, static configuration to a dark radiation and dark pressure)
modifies space-time geometry around galaxies and has been shown to explain
the flatness of galactic rotation curves. Independent observations for
discerning between the Weyl fluid and other dark matter models are
necessary. Gravitational lensing could provide such a test. Therefore we
study null geodesics and weak gravitational lensing in the dark radiation
dominated region of galaxies in a class of spherically symmetric brane-world
metrics. We find that the lensing profile in the brane-world scenario is
distinguishable from dark matter lensing, despite both the brane-world
scenario and dark matter models fitting the rotation curve data. In
particular, in the asymptotic regions light deflection is $18\%$ enhanced as
compared to dark matter halo predictions. For a linear equation of state of
the Weyl fluid we further find a critical radius, below which brane-world
effects reduce, while above it they amplify light deflection. This is in
contrast to any dark matter model, the addition of which always increases
the deflection angle.
\end{abstract}

\date{\today}
\maketitle

\section{Introduction}

The idea of embedding our Universe in a higher dimensional non-compactified
space-time has attracted considerable interest in the last decade, due to
the proposal by Randall and Sundrum ~\cite{Randall} that our
four-dimensional (4d) space-time could be a three-brane embedded in a
5-dimensional (5d) space-time (the bulk). According to the brane-world
scenario, the physical fields (electromagnetic, Yang-Mills etc.) observed in
our 4d Universe are confined to the three-brane. Only gravity can freely
propagate both on the brane and in the bulk, with the gravitational
self-couplings not significantly modified. The model allows for a large, or
even infinite non-compact extra dimension, in the simplest case the brane
being identified to a domain wall in a 5d anti-de Sitter space-time. Even
with the fifth dimension uncompactified, standard 4d gravity can be
reproduced on the brane in a certain limit. For a review of the dynamics and
geometry of brane universes, see e.g.~\cite{Mar04}.

At very high energies, in the presence of large 5d
curvatures, significant deviations from the standard Einstein theory could occur in
brane-world models, due to the nonstandard model 5d fields, possible asymmetric
embeddings of the brane into the bulk or a variable brane tension~\cite%
{SMS00,311}. At the electro-weak scale of about
1~TeV gravity is largely modified. The cosmological and astrophysical implications of the brane-world
theories have been extensively investigated in the physical literature~\cite%
{all2,HGH,sol}.

On the vacuum brane the gravitational field equations  depend on the
generally unknown brane projections of the Weyl curvature of the bulk,
generating non-local brane stresses, which in a spherically symmetric setup
can be expressed in terms of two functions, the dark radiation $U$ and the
dark pressure $P$, respectively \cite{Da00,Mar04,GeMa01}. On a vacuum
brane several classes of spherically symmetric solutions of the static gravitational field equations have been found in \cite{Ha03,Ma04,Ha05}.

Dark matter is one of the central issues in modern astrophysics (see \cite{Sal1} for an extensive review of the recent results of the search for dark matter). The necessity of considering the existence of dark matter at a galactic and
extra-galactic scale is required by two fundamental  observational evidences, the behavior of the galactic rotation
curves, and the mass discrepancy in clusters of galaxies, respectively. On the
galactic/intergalactic scale the rotation curves of spiral galaxies~\cite%
{Binney, Sal2, Sal3} provide one of the best evidences showing the problems Newtonian
gravity and/or standard general relativity have to face. The rotational velocities of hydrogen clouds
in stable circular orbits increase near the center of the galaxy, in
agreement with the standard gravitational theory, but then remain nearly constant at a value of $%
v_{tg\infty }\sim 200\div 300$ km/s~\cite{Binney}. Hence we obtain a mass
profile of the form $M(r)=rv_{tg\infty }^{2}/G$. This result implies that the mass
within a distance $r$ from the center of the galaxy increases linearly with $%
r$, even at large distances where very little luminous matter does exist.

The second important astrophysical evidence for dark matter comes from the study of
the clusters of galaxies. Generally it is found that the
virial mass $M_{VT}$ is considerably greater than the observed baryonic mass
$M_{B}$, $M_{VT}>M_{B}$, typical values of $M_{VT}/M_{B}$ being about 20 - 30~%
\cite{Binney}.

This behavior of the galactic rotation curves and of the virial mass of
galaxy clusters is usually explained by postulating the existence of some
dark (invisible) matter, distributed in a spherical halo around the
galaxies. The dark matter is assumed to be a cold, pressure-less medium.
Many possible candidates for dark matter have been proposed, the most popular ones
being the  WIMPs (Weakly Interacting Massive Particles) (for a review of the
particle physics aspects of the dark matter see~\cite{OvWe04}). While extremely
small, their interaction cross sections with normal baryonic matter, are expected to be non-zero, so that their direct experimental detection may be possible.

However, up to now no direct evidence or annihilation radiation from dark matter has been
observed, and no non-gravitational evidence for dark matter does exist. Therefore, it seems that the possibility that Einstein's (and the Newtonian) theory of gravity breaks down at the scale of galaxies cannot
be excluded \textit{a priori}.

In brane-world models the rotational galactic curves can be naturally explained without introducing any supplementary hypothesis \cite{Ma04,Ha05}. The non-zero contribution of the Weyl tensor from the bulk
 generates a modified, spherically symmetric geometry, in which galaxies are embedded. The dark
radiation $U$ and the dark pressure $P$ act similarly to a \textquotedblleft
matter\textquotedblright\ distribution outside the galaxy. The particles
moving in this geometry feel the gravitational effects of $U$, which can be
expressed in terms of an equivalent mass (the dark mass) $M_{U}$. The dark
mass is linearly increasing with the distance, and proportional to the
baryonic mass of the galaxy, $M_{U}(r)\approx M_{B}(r/r_{0})$~\cite{Ma04}.

Therefore it would be of uttermost importance to have independent tests,
which could discern between various dark matter models and modified gravity
models, which include brane-worlds.

Gravitational lensing and the study of the light deflection by black holes
and galaxies is an important physical effect that could provide specific
signatures for testing the brane-world models (for a review of the
gravitational lensing by brane-world black holes see \cite{Majum05}).
Observables related to the relativistic images of strong field gravitational
lensing could in principle be used to distinguish between different
brane-world black hole metrics in future observations.

It is the purpose of the present paper to consider the lensing in the dark
radiation dominated region of the brane-world models. Physically, this region corresponds to particles gravitating in circular orbits and at constant speed around the galactic center~\cite{Binney}.

The galactic rotation curves show a large variety of behaviors, and, in particular, they also depend on the considered galaxy type. By analyzing an extended set of spiral galaxy rotation curves it has been proposed that the rotation curves of these galaxies can be described by a Universal Rotation Curve (URC) \cite{URC}. N-body simulations provide a universal mass profile in the $\Lambda $ Cold Dark Matter ($\Lambda $CDM) cosmological scenario, and, consequently, a universal equilibrium circular velocity of the virialized objects, as galaxies. By combining kinematical data of inner galactic regions with global observational properties, the URC of disc galaxies and the corresponding mass distribution out to their virial radius was obtained. The existence of a Universal Rotation Curve is also consistent with the predictions of the brane world models, which imply the existence of of a Universal Weyl Fluid, acting in a uniform way at all galactic scales. In the present paper,  we will analyze the lensing properties in both the constant velocity region of the rotation curves, as well as in the declining or increasing regions, where the behavior of the rotation curves can be modeled by a simple power law \cite{Sal1}.

By fixing two radius values at the observer and source, respectively, we
derive an exact lens equation relating two angular variables. This equation
allows to obtain all the observationally relevant quantities, like image
position, the apparent brightness and the image distortions.

The present paper is organized as follows. We review the basic properties of the brane world models and of the
Weyl fluid in Section~\ref{sect2}. The metric properties valid in the region of constant
or power law type tangential velocities are also discussed. The mathematical problem of the
embedding is considered in Section~\ref{emb}. Particular solutions of the field equations describing the geoemtric properties in the constant and power law velocity regions are
presented in Section~\ref{sect4}. The deflection of light and the lensing properties in the brane world models in the Weyl fluid dominated regions are obtained in Section~\ref{sect5}. In Section~\ref{sect6} we compare the theoretical predictions of the model with the observational data.  We discuss and conclude our results in
Section~\ref{sect7}.

\section{The Weyl fluid in a spherically symmetric, static vacuum brane}\label{sect2}

\subsection{A succinct introduction to brane-worlds}

In the brane-world model standard model fields are confined to the 4d brane,
a hypersurface $({}M,g_{\mu \nu })$ embedded (for simplicity, in a $Z_{2}$
symmetric way)\ in a 5d bulk space-time $({}^{(5)}M,^{(5)}g_{AB})$. We
denote the coordinates in the bulk and on the brane as $X^{A},A=0,1,...,4$
and $x^{\mu },\mu =0,1,2,3$, respectively. The system is characterized by
the combined action~\cite{SMS00}
\begin{equation}
S=S_{bulk}+S_{brane},  \label{bulk}
\end{equation}%
with
\begin{equation}
S_{bulk}=\int_{{}^{(5)}M}\sqrt{-{}^{(5)}g}\left[ \frac{1}{2k_{5}^{2}}{}%
^{(5)}R+{}\Lambda _{5}\right] d^{5}X,
\end{equation}%
and
\begin{equation}
S_{brane}=\int_{{}M}\sqrt{-{}g}\left[ \frac{1}{k_{5}^{2}}K^{\pm
}+L_{brane}\left( g_{\alpha \beta },\psi \right) +\lambda _{b}\right] d^{4}x,
\end{equation}%
where $k_{5}^{2}=8\pi G_{5}$ is the 5d gravitational constant, ${}^{(5)}R$
is the 5d scalar curvature, $\Lambda _{5}$ is the 5d cosmological constant.
In the boundary terms $K^{\pm }$ denote the traces of the extrinsic
curvatures taken on the two sides of the brane, $L_{brane}\left( g_{\alpha
\beta },\psi \right) $ is the 4d matter Lagrangian, a generic functional of
the brane metric $g_{\alpha \beta }$ and standard model fields $\psi $,
finally $\lambda _{b}$ is the brane tension (chosen here for simplicity a
constant).

The Einstein field equations in the bulk emerge as~\cite{SMS00}
\begin{equation}
{}^{(5)}G_{IJ}=-\Lambda _{5}{}^{(5)}g_{IJ}+\delta (0)\left[ -\lambda
_{b}{}^{(5)}g_{IJ}+T_{IJ}\right] ,  \label{5dEinstein}
\end{equation}%
where $g_{IJ}$ is the induced brane metric expressed in bulk coordinates,
the Dirac delta-function $\delta \left( 0\right) $ appears in the source
terms localized on the brane, including the brane tension and $T_{IJ}$, the
energy-momentum tensor of the standard model fields, defined in brane
coordinates as
\begin{equation}
T_{\mu \nu }\equiv -2\frac{\delta L_{brane}}{\delta g^{\mu \nu }}+g_{\mu \nu
}\text{ }L_{brane}.
\end{equation}%
The brane energy momentum tensor is related to the jump in the extrinsic
curvature $K_{\mu \nu }=\mathcal{L}_{n}g_{\mu \nu }$ by the Lancyos equation
(the second Israel junction condition)
\begin{equation}
K_{\mu \nu }(x^{\mu },y=+0)-K_{\mu \nu }(x^{\mu },y=-0)
=-k_{5}^{2}\left( T_{\mu \nu }-\frac{1}{3}g_{|mu\nu }T\right) .
\end{equation}%
The 4d equations emerge as projections to the brane with the induced metric $%
g_{IJ}$ and contractions with the brane unit normal $n^{I}$ (the brane has
the normal form $n_{I}dx^{I}=dy$ and is chosen at $y=0$). The tensorial
projection is the effective Einstein equation \cite{SMS00}

\begin{equation}
G_{\mu \nu }=-\Lambda g_{\mu \nu }+k_{4}^{2}T_{\mu \nu }+k_{5}^{4}S_{\mu \nu
}-E_{\mu \nu },  \label{Ein}
\end{equation}%
with $S_{\mu \nu }$ the quadratic energy-momentum correction (to be ignored
at infrared scales)
\begin{equation}
S_{\mu \nu }=\frac{1}{12}TT_{\mu \nu }-\frac{1}{4}T_{\mu }{}^{\alpha }T_{\nu
\alpha }+\frac{1}{24}g_{\mu \nu }\left( 3T^{\alpha \beta }T_{\alpha \beta
}-T^{2}\right) ,
\end{equation}%
and $E_{\mu \nu }$ a non-local effect from the bulk gravitational field, the
electric projection $E_{IJ}=C_{IAJB}n^{A}n^{B}$ of the bulk Weyl tensor $%
C_{IAJB}$ (with $E_{IJ}\rightarrow E_{\mu \nu }\delta _{I}^{\mu }\delta
_{J}^{\nu }\quad $as$\quad y\rightarrow 0$). We denote $k_{4}^{2}=8\pi G$,
with $G$ the 4d gravitational constant. The 4d cosmological constant $%
\Lambda $, and the 4d coupling constant $k_{4}$ are related to the brane
tension as $\Lambda =k_{5}^{2}(\Lambda _{5}+k_{5}^{2}\lambda _{b}^{2}/6)/2$
and $k_{4}^{2}=k_{5}^{4}\lambda _{b}/6$, respectively.

The Codazzi equation
\begin{equation}
\nabla _{B}K_{A}^{B}-\nabla _{A}K={}^{(5)}R_{BC}g_{A}^{B}n^{C}
\end{equation}%
implies the conservation of the energy-momentum tensor of the matter on the
brane, $D_{\nu }T_{\mu }{}^{\nu }=0$, where $D_{\nu }$ denotes the brane
covariant derivative. Moreover, from the contracted Bianchi identities on
the brane it follows that the projected Weyl tensor obeys the constraint
\begin{equation}\label{WeylC}
D_{\nu }E_{\mu }{}^{\nu}=k_{5}^{4}D_{\nu }S_{\mu }{}^{\nu }.
\end{equation}
For low density region, we have,
\begin{equation}
D_{\nu }E_{\mu }{}^{\nu }=0.  \label{Ecd}
\end{equation}

\subsection{The Weyl fluid}

The 5D vacuum metric containing a spherical symmetric, static, vacuum brane
could be written in terms of normal coordinates as
\begin{equation}\label{5dmetric}
{}^{(5)}ds^{2}=g_{AB}dx^{A}dx^{B}=-M(r,y)^{2}dt^{2}+N(r,y)^{2}dr^{2}
+Q(r,y)^{2}\left( d\theta ^{2}+\sin \theta ^{2}d\phi ^{2}\right) +dy^{2}.
\end{equation}%
The corresponding electric part of the Weyl curvature becomes
\begin{equation}
E_{\nu }^{\mu }=diag\left( -\frac{\Lambda _{5}}{2}+\frac{2\widetilde{Q_{,y,y}%
}}{Q}+\frac{2\widetilde{N_{,y,y}}}{N},\right.  \label{Ebulk}
\left. \frac{\Lambda _{5}}{6}-\frac{\widetilde{N_{,y,y}}}{N},\frac{\Lambda
_{5}}{6}-\frac{\widetilde{Q_{,y,y}}}{Q},\frac{\Lambda _{5}}{6}-\frac{%
\widetilde{Q_{,y,y}}}{Q}\right) ,
\end{equation}%
where we followed the decomposition \cite{Binetruy}
\begin{eqnarray}
N_{,y,y} &=&\delta (y)N_{,y}+\widetilde{N_{,y,y}},  \label{FDdecomposition}
\\
Q_{,y,y} &=&\delta (y)N_{,y}+\widetilde{Q_{,y,y}},
\end{eqnarray}%
into a distributional part and an analytic function (regular) part. The
regular part is subject to the constraint~(\ref{Ecd}), characterizing the
bulk effects on the brane. On the other hand $\widetilde{N_{,y,y}}$ and $%
\widetilde{Q_{,y,y}}$ depend on the embedding of the brane, they
corresponding to the jump in $N_{,y}$ and $Q_{,y}$, respectively, which
arise from the jump in the extrinsic curvature.

The symmetry properties of $E_{\mu \nu }$ also imply the irreducible
decomposition with respect to a chosen $4$-velocity field $u^{\mu }$ on the
brane~\cite{Mar04}
\begin{equation}
E_{\mu \nu }=-k^{4}\left[ U\left( u_{\mu }u_{\nu }+\frac{1}{3}h_{\mu \nu
}\right) +P_{\mu \nu }+2V_{(\mu }u_{\nu )}\right] ,  \label{WT}
\end{equation}%
where $k=k_{5}/k_{4}$, $h_{\mu \nu }=g_{\mu \nu }+u_{\mu }u_{\nu }$ projects
orthogonal to $u^{\mu }$, the \textit{dark radiation} $U=-k^{4}E_{\mu \nu
}u^{\mu }u^{\nu }$ is a scalar, $V_{\mu }=k^{4}h_{\mu }^{\alpha }E_{\alpha
\beta }$ is a spatial vector and
\begin{equation}
P_{\mu \nu }=-k^{4}\left[ h_{(\mu }\text{ }%
^{\alpha }h_{\nu )}\text{ }^{\beta }-\frac{1}{3}h_{\mu \nu }h^{\alpha \beta }%
\right] E_{\alpha \beta },
\end{equation}
is a spatial, symmetric and trace-free tensor.

In a static spherical spacetime $V_{\mu }=0$, and Eq. (\ref{Ecd}) takes the
explicit form \cite{GeMa01}:
\begin{equation}
\frac{1}{3}D_{\mu }U+\frac{4}{3}UA_{\mu }+D^{\nu }P_{\mu \nu }+A^{\nu
}P_{\mu \nu }=0,
\end{equation}%
with $A_{\mu }=u^{\nu }D_{\nu }u_{\mu }$ the 4-acceleration. By further
assuming spherical symmetry, we may chose $A_{\mu }=A(r)r_{\mu }$ and
\begin{equation}
P_{\mu \nu }=P(r)\left( r_{\mu }r_{\nu }-\frac{1}{3}h_{\mu \nu }\right) ,
\end{equation}
where $P(r)$ is the \textit{dark pressure} (although the term dark
anisotropic stress might be more appropriate) and~$r_{\mu }$ is a unit
radial vector~\cite{Da00}.

Characterized by just two functions $U$ and $P\,$, the electric part $E_{\mu
\nu }$ of the Weyl curvature can be regarded as a perfect fluid of geometric
origin, the source of gravity on the brane. We call this "fluid" the Weyl
fluid.

The Weyl fluid is a macroscopic fluid with an effective energy density and pressure, similar to any other material fluid. However, the energy and the pressure of the fluid depend and are determined by the geometry of the extra dimension. Moreover, it satisfies a more complicated continuity equation than usual material fluids, Eq.~(\ref{WeylC}), and it is traceless. If we observe a Weyl -type fluid behavior in the Universe, the most natural explanation will be the brane world effects. Using the hypothesis of the Weyl fluid as a solution for the missing mass problem is a different approach than assuming the existence of a new particle, generating a standard material fluid in general relativistic framework.

\subsection{Dark matter as a bulk effect}

In this subsection we discuss a brane metric properly reproducing the
galactic rotation curves. We assume that galactic dynamics is governed by
the static and spherically symmetric 4d line element
\begin{equation}
ds^{2}=-e^{\nu (r)}dt^{2}+e^{\lambda (r)}dr^{2}+r^{2}d\Omega ^{2},
\label{metr1}
\end{equation}%
where $d\Omega ^{2}=d\theta ^{2}+\sin ^{2}\theta d\phi ^{2}$. Here $\theta $
and $\phi $ are spherical coordinates, $t\in R$ and $r$ ranges over an open
interval $\left( r_{\min },r_{\max }\right) $ so that $0\leq r_{\min }\leq
r_{\max }\leq \infty $. We also assume that the functions $\nu (r)$ and $%
\lambda (r)$ are strictly positive and (at least piecewise) differentiable
on the interval $\left( r_{\min },r_{\max }\right) $.

In the metric (\ref{metr1}) the independent components of the effective
Einstein equation are
\begin{eqnarray}
&&-e^{-\lambda }\left( \frac{1}{r^{2}}-\frac{\lambda ^{\prime }}{r}\right) +%
\frac{1}{r^{2}}=-E_{t}^{t},  \label{field1} \\
&&e^{-\lambda }\left( \frac{\nu ^{\prime }}{r}+\frac{1}{r^{2}}\right) -\frac{%
1}{r^{2}}=-E_{r}^{r},  \label{field2} \\
&&\frac{e^{-\lambda }}{2}\left( \nu ^{\prime \prime }+\frac{\nu ^{\prime 2}}{%
2}+\frac{\nu ^{\prime }-\alpha ^{\prime }}{r}-\frac{\nu ^{\prime }\alpha
^{\prime }}{2}\right) =-E_{\theta }^{\theta },  \notag  \label{field3} \\
&&
\end{eqnarray}%
where a prime denotes the derivative with respect to $r$. Knowing the brane
metric, the projected Weyl tensor is determined by this system.

The tangential velocity $v_{tg}$ of a test particle is measured in terms of
proper time, that is, by an observer located at the given point, as~\cite%
{Matos, Nuc01}
\begin{equation}
v_{tg}^{2}=e^{-\nu }r^{2}\dot{\Omega}^{2}.  \label{vtgbr}
\end{equation}%
By using the constants of motion, the tangential velocity of a test particle
in a stable circular orbit on the brane emerges as~\cite{Matos, Nuc01}:
\begin{equation}
v_{tg}^{2}=\frac{r\nu ^{\prime }}{2}.  \label{vtg}
\end{equation}%
In a power law tangential velocity limiting case \cite{Sal1},
\begin{equation}
v_{tg}(r)\to v_c r^{\zeta },
\end{equation}
where $v_c$ and $\zeta$ are constants.

This power law velocity profile could be the limiting case of various velocity profile. For example, such a behavior is suggested by the Universal Rotation Curve model \cite{Sal2, URC}. The corresponding metric coefficient is given by
\begin{equation}\label{limitc}
\nu=\left\{
\begin{array}{ll}
\frac{v_c^2 r^{2\zeta}}{\zeta }+\nu_0, &\mbox{for $\zeta \neq0$ },\\
\ln\left(\frac{r}{R_{\infty}}\right)^{2v_{c}^2}, &\mbox{for $\zeta =0$.},
\end{array}
\right.
\end{equation}
where $\nu_0$ and $R_{\infty }$ are arbitrary constants of integration.

In particular, constant tangential velocity limit immediately allows us to find the metric tensor component $e^{\nu
}$ in the
flat rotation curves region on the brane as
\begin{equation}
e^{\nu }=\left( \frac{r}{R_{\infty }}\right) ^{2v_{tg}^{2}}.  \label{nu}
\end{equation}%

Since $E_{\mu\nu}$ is traceless, summing all Eq.~(\ref{field1})-Eq.~(\ref{field3}) could determine
\begin{equation}
e^{-\lambda }\approx 1-v_{tg}^{2}.  \label{lam}
\end{equation}%
Eqs.~(\ref{nu}).and~(\ref{lam}) give the galactic metric in the constant
velocity region on the brane.

The brane metric obtained phenomenologically by the requirement that the
required tangential velocities are realized in the model could furter be
specified by assuming a particular embedding of the brane in the bulk,
equivalent with the equation of state of the Weyl fluid
\begin{equation}
P=\left( a-2\right) U-\frac{B}{3\alpha _{b}r^{2}}~.  \label{eqstate}
\end{equation}%
Here $3\alpha _{b}=64\pi ^{2}G^{2}$ and $B$ is an arbitrary constant of
integration. It has been shown, that this choice of the equation of state
leads to a tangential velocity compatible with the observations on rotation
curves \cite{Ma04, BraneRotCurves}. An explicit proof based on the 3+1+1
decomposed covariant bulk dynamics \cite{311} has shown that the
corresponding five dimensional embedding does exists \cite{BraneRotCurves}.

In the next Section we will discuss  the embedding of this brane metric into
5d.

\section{The embedding problem}\label{emb}

In the previous Section we have used the observed characteristics of the rotation
curves to fix $e^{\nu }$, and the traceless property of the electric
part of the Weyl tensor to fix $e^{\lambda }$. We need to prove
that our choices are compatible with an integral bulk solution, i.e. the
derived phenomenological metric can be realized on a brane embedded in some
bulk geometry satisfying the 5d Einstein equations (\ref{5dEinstein}).

\subsection{The Campbell-Magaard theorem}

The Campbell-Magaard theorem states that it is possible to embed a manifold
with arbitrary geometry into a Ricci-flat manifold with one extra dimension
\cite{Campbell, Magaard}. The use of the Campbell-Magaard theorem for the
embedding problem in higher dimensional theories has been studied recently
\cite{Dahia1, Chervon, Dahia2, Leon}. In \cite{Seahra} it was shown that it is possible
to generalize the Campbell-Magaard theorem to the Randall-Sundrum type
brane-world model, as any solution of 4d projected field
equation can be realized as a thin 3-brane in the brane-world. However, if
one considers the 5d geodesic corrections to particle dynamics, the matter
on brane will leak off if the total energy momentum on the brane (including
the brane tension) does not satisfy the strong energy condition. Although we
do not consider higher dimensional particle motion, we include a discussion
on the embedding based on the Campbell-Magaard theorem. We find that a
metric reproducing rotation curve dynamics could be realized in the
Randall-Sundrum type brane-world, i.e. there is a 5d spacetime ($M$,$g_{MN}$%
) satisfying the field equations such that the brane is a representant of a
foliation of $M$ with codimension 1, giving the rotation curve dynamics. In
the following subsection, we will give the details based on the procedure
described in \cite{Seahra}:

\begin{enumerate}
\item Write down a brane metric $g_{\mu \nu }$, extrinsic curvature $K_{\mu \nu }$
and projected Weyl tensor $E_{\mu \nu }$ that is consistent with the
equations for $g_{MN}$ evaluated on the brane.

\item Show the existence of a unique solution for $g_{MN}$ on the half plane
$y>0$ with boundary condition given in 1.

\item Show the existence of another unique solution for equations of $g_{MN}$ on half plane $%
y<0$.

\item The union of the two solution will be our bulk spacetime.
\end{enumerate}

However, the bulk geometry will not be AdS5, and thus gravity proves to be
localized on the brane, with a Newtonian limit. This is acceptable, as there
is no independent experimental test for gravity and particle physics on the
galactic scale. In the following subsection we discuss the details of this
procedure.

\subsection{Existence of the bulk solution}

We will make use of Cauchy-Kowalewski theorem for non-linear partial
differential equation (PDE) \cite{FRIEDMAN}. The theorem states that a
unique analytic solution exist in the neighborhood of $y=0$ for an analytic
Cauchy problem. The analytic Cauchy problem is defined by a system of PDEs
with a Cauchy data. The PDEs have the form
\begin{equation}
\partial _{y}^{m}u=F_{i}(x,y,u(x,y),u^{(\alpha )})
\end{equation}%
where $x$ can be an array, $u^{(\alpha )}$ is the derivative of the dependent
variable $u$ to order $\alpha \leq m$, and $F$ is analytic. The Cauchy data
are a set of boundary conditions at $y=0$,
\begin{equation}
\partial _{y}^{k}u=f^{k};\mbox{ }0\leq k<m
\end{equation}%
where $f^{k}$ is analytic for every $k<m$.

\subsection{The Cauchy problem in the braneworld model}

We are going to translate the braneworld embedding problem into a Cauchy
type problem. In the brane world model we are considering a weak solution of
the Einstein equations, i.e., $g_{MN}(x^{\mu },y)$ satisfies the 5D Einstein
equation for a region $y\neq 0$, and the brane is a discontinuity
on $\partial _{y}g_{MN}(x^{\mu },y)$ at $y=0$. Therefore, we first look at
the solution of the 5D Einstein equation in $y\in \lbrack 0,+\infty )$.

On the other hand, using the function-distribution decomposition, Eq.~(\ref%
{FDdecomposition}), the 5D Einstein equations become PDEs without delta
functions that are valid on each half plane.
\begin{eqnarray}
{}^{(5)}G_{t}^{t} &=&-\frac{2Q_{,r}N_{,r}}{QN^{3}}+\frac{2Q_{,y}N_{,y}}{NQ}+%
\frac{2Q_{,r,r}}{QN^{2}}+\frac{N_{,y,y}}{N}  \notag  \label{analyticE1} \\
&+&\frac{2Q_{,y,y}}{Q}+\Psi =\Lambda _{5},
\end{eqnarray}%
\begin{eqnarray}
{}^{(5)}G_{r}^{r} &=&\frac{2Q_{,r}M_{,r}}{MQN^{2}}+\frac{2Q_{,y}M_{,y}}{MQ}+%
\frac{M_{,y,y}}{M}  \notag  \label{analyticE2} \\
&+&\frac{2Q_{,y,y}}{Q}+\Psi =\Lambda _{5},
\end{eqnarray}%
\begin{equation*}
{}^{(5)}G_{r}^{y}=\left( \frac{M_{,r}}{MN}+\frac{2Q_{,r}}{NQ}\right) -\frac{%
M_{,r,y}}{M}-\frac{2Q_{,r,y}}{Q}=0,
\end{equation*}%
\begin{eqnarray}
{}^{(5)}G_{\theta }^{\theta } &=&\left( \frac{Q_{,y}}{Q}+\frac{N_{,y}}{N}%
\right) \frac{M_{,y}}{M}+\frac{Q_{,y}N_{,y}}{QN}-\frac{Q_{,r}N_{,r}}{QN^{3}}
\notag  \label{analyticE4} \\
&+&\left( \frac{Q_{,r}}{Q}-\frac{N_{,r}}{N}\right) \frac{M_{,r}}{MN^{2}}+%
\frac{Q_{,y,y}}{Q}+\frac{Q_{,r,r}}{N^{2}Q}  \notag \\
&+&\frac{N_{,y,y}}{N}+\frac{M_{,r,r}}{MN^{2}}+\frac{M_{,y,y}}{M}=\Lambda
_{5},
\end{eqnarray}%
\begin{eqnarray}
{}^{(5)}G_{y}^{y} &=&\left( \frac{2Q_{,y}}{Q}+\frac{N_{,y}}{N}\right) \frac{%
M_{,y}}{M}+\frac{2Q_{,r,r}}{N^{2}Q}+\frac{M_{,r,r}}{MN^{2}}  \notag
\label{analyticE5} \\
&+&\left( \frac{2Q_{,r}}{Q}-\frac{N_{,r}}{N}\right) \frac{M_{,r}}{MN^{2}}-%
\frac{2Q_{,r}N_{,r}}{N^{3}Q}  \notag \\
&+&\frac{2Q_{,y}N_{,y}}{NQ}+\Psi =0,
\end{eqnarray}%
where
\begin{equation}
\Psi =\frac{Q_{,r}^{2}}{Q^{2}N^{2}}-\frac{1}{Q^{2}}+\frac{Q_{,y}^{2}}{%
Q^{2}}.
\end{equation}
 This system forms the analytic PDEs of the Cauchy problem.

\subsection{Analyticity and consistency of Cauchy data}

We need to show that the way we fixed the 4d metric in our previous Section
gives consistent Cauchy data. We determine $e^{\nu }$ by the rotation curve
and determine $e^{\lambda }$ by $TrG_{\mu \nu }=0$. On the other hand the
Cauchy data relevant to Eqs.~(\ref{analyticE1})-(\ref{analyticE5}) consist
of $g(x^{\mu },0)$ and $\partial _{y}g(x^{\mu },0)$.

The dynamics inside galaxies, says for $r<r_{c}$, is not described by our brane
metric. The metric specified by Eqs.~(\ref{metr1}), (\ref{vtgbr}) and the
traceless property of $E_{\mu \nu }$ only applies to the outer region of
galaxies, so that they have a domain of $r\in \lbrack r_{c},\infty )$. Therefore
the physical situation under consideration isolates singularities that may
arise in the brane metric, i.e., fixing the 4d metric determines $g(x^{\mu
},0) $, which is analytic on $r\in (r_{c},\infty )$.

On the other hand, the junction condition determines $\partial _{y}g(x^{\mu
},+0)$, which is analytic. However, $g(x^{\mu },0)$ also tells about $%
\partial _{yy}^{2}g(x^{\mu },0)$ by equating Eq.~(\ref{Ebulk}) with Eq.~(\ref%
{field1}) - Eq.~(\ref{field3}). But these equations must be consistent with
Eqs.~(\ref{analyticE1})-(\ref{analyticE5}) as Eqs.~(\ref{Ebulk}) are derived
from their geometric projection. One could also verify these by inserting
them into Eqs.~(\ref{analyticE1})-(\ref{analyticE5}).

\subsection{$Z_2$ symmetry and $AdS_5$ bulk}

We know that a unique solution on the half plane $y\in \lbrack 0,+\infty )$
does exists. By putting $g(x^{\mu },z)=g(x^{\mu },-y)$ into Eqs.~(\ref%
{analyticE1})-(\ref{analyticE5}), we obtain the PDEs on the another half
plane. $g(x^{\mu },0)$ and $\partial _{y}g(x^{\mu },z=+0)$ give the Cauchy
data for the another half plane. The uniqueness of solution, the symmetry of
Eqs.~(\ref{analyticE1})-(\ref{analyticE5}), and the Cauchy data shows that
the solution on the another half plane is identical, i.e. $g(x^{\mu
},-y)=g(x^{\mu },y)$.

However, there is no guarantee of an asymptotic AdS5 solution. Such a
requirement would over-determine the Cauchy problem. This can be seen, for
example, from the numerical analysis of the spherically symmetric solution
\cite{Wiseman}. A numerical solution for 5d metric that is ADS
asymptotically can be obtained without any phenomenological assumption on
the brane metric. As we have discussed in the Introduction section,
experimental confirmation of Newtonian gravity on galactic scale is still
lacking.

\section{Particular solutions of the field equations in the Weyl fluid
dominated region}\label{sect4}

In the present Section we present particular solutions of the gravitational
field equations on the brane in the $E_{\mu \nu }$ dominated regions (where
we include baryonic matter only as test particles).

The metric that describes the spacetime in the spherically symmetric galactic
halo region, with the baryonic matter concentrated in the central part of the
galaxy, is the Schwarzschild solution of general relativity. A brane-world
solution could emerge as a correction to the Schwarzschild metric, with
\begin{eqnarray}
e^{\nu (r)} &=&1-\frac{r_{S}}{r}+B_{t}(r)~,  \label{generic4d} \\
e^{\lambda (r)} &=&\frac{1}{1-\frac{r_{S}}{r}+B_{r}(r)}~.
\end{eqnarray}%
Here $r_{S}$ is the Schwarzschild radius. The functions $B_{t}(r)$ and $%
B_{r}(r)$ should be compatible with a traceless Einstein tensor $G_{\mu \nu }
$ in order to be realizable in Randall Sundrum brane-world model. The
traceless condition gives rise to the following equation%
\begin{equation}\label{traceless}
TrG=\frac{B_{t}^{\prime \prime }(r-r_{S}+rB_{r})}{r-r_{S}+rB_{t}}-\frac{%
rB_{t}^{\prime 2}(rB_{r}+r-r_{S})}{2(r-r_{S}+rB_{r})}
+\frac{rB_{r}^{\prime }B_{t}^{\prime }(r-r_{S}+rB_{t})}{2(r-r_{S}+rB_{t})}.
\end{equation}

The question comes whether it is possible to have a solution for $B_{t}$, $%
B_{r}$ consistent with the mass required from galactic rotation curves. For
this, $B_{t}(r)$ should be related to the tangential velocity profile $%
v_{tg}(r)$ of the galaxies through
\begin{equation}
v_{tg}^{2}(r)=\frac{r\nu ^{\prime }}{2}.
\end{equation}%
Substituting this equation into the metric (\ref{generic4d}), we obtain the differential
equation for $B_{t}$ in terms of the tangential velocity profile
\begin{equation}
B_{t}^{\prime }-\frac{2v_{tg}^{2}}{r}\left( 1-\frac{r_{S}}{r}+B_{t}\right) +%
\frac{r_{S}}{r^{2}}=0.  \label{DEBt}
\end{equation}%
Eliminating $B_{t}^{\prime }$ from Eq.~(\ref{traceless}) leads to a
differential equation for $B_{r}$,
\begin{equation}\label{DEBr}
(2+v_{tg}^{2})B_{r}^{\prime }+\left( \frac{2}{r}+4v_{tg}v_{tg}^{\prime }+%
\frac{2v_{tg}^{2}}{4}+\frac{2v_{tg}^{4}}{r}\right) B_{r}
+v_{tg}^{4}\left( \frac{2}{r}-\frac{2r_{S}}{r^{2}}\right) +v_{tg}^{2}\left(
\frac{2}{r}-\frac{r_{S}}{r^{2}}\right) +4v_{tg}v_{tg}^{\prime }\left( 1-%
\frac{r_{S}}{r}\right) =0.
\end{equation}

The solution of this system of ordinary differential equation gives the
corrected Schwarzschild type solution of the Einstein equations.

\subsection{The metric in the flat rotation curves region}

In the limit of constant tangential velocity, i.e. $v_{tg}\rightarrow
v_{\infty }$, Eq.~(\ref{DEBt}) can be integrated and leads to the asymptotic
behavior
\begin{equation}
B_{t}(r)\rightarrow -1+\frac{r_{S}}{r}+C_{t}r^{2v_{\infty }^{2}}.
\label{asymBt}
\end{equation}%
Eq.~(\ref{DEBr}) can be integrated to give

\begin{equation}\label{asymBr}
B_{r}(r)\rightarrow r^{-\frac{2v_{\infty }^{4}+2v_{\infty }^{2}+2}{v_{\infty
}^{2}+2}}\left[ C_{r}-\right.
\left. r^{\frac{v_{\infty }^{2}(1+2v_{\infty }^{2})}{v_{\infty }^{2}+2}%
}\left( \frac{(v_{\infty }^{4}+v_{\infty }^{2})r}{v_{\infty }^{4}+v_{\infty
}^{2}+1}-r_{S}\right) \right] ,
\end{equation}%
with $C_{t}$, $C_{r}$ being integration constants.

The virial mass is derived from the observed acceleration of massive
baryonic particles, for example, in X-ray cluster images. It will be useful
to see the acceleration of massive particles in the weak field, low velocity
limit of our metric. Massive particles move on brane geodesics
\begin{equation}
\frac{d^{2}x^{\mu }}{d\tau ^{2}}+\Gamma _{\rho \sigma }^{\mu }\frac{dx^{\rho
}}{d\tau }\frac{dx^{\sigma }}{d\tau }=0~,
\end{equation}%
with $\tau $ the proper time. At small velocities $dx^{\mu }/d\tau \approx
(1,0,0,0)$ and $t\approx \tau $ hold \cite{Wald}, and thus
\begin{equation}
\frac{d^{2}x^{\mu }}{dt^{2}}=-\Gamma _{tt}^{\mu }~,
\end{equation}%
\begin{equation}
a=\frac{1}{2r^{5}}\left( 1-\frac{r_{S}}{r}+B_{r}\right) \left( \frac{r_{S}}{r%
}+rB_{t}^{\prime }\right) ~.
\end{equation}%
Inserting the expressions for $B_{t}$, $B_{r}$ from Eqs.~(\ref{asymBt}), (%
\ref{asymBr}) and keeping the leading order terms in the small parameters $%
r_{S}/r$ and $v_{\infty }^{2}$ gives
\begin{equation}
a\approx \frac{v_{\infty }^{2}}{r}
\end{equation}%
The acceleration derived from this brane-world metric agrees with the
Newtonian gravitational acceleration of dark matter halo with inverse square
law radial dependence of the gravitational force.

\subsection{The unified Schwarzschild - constant velocity metric}

We consider here a tangential velocity arising from a baryonic central mass
and some unknown mass as
\begin{equation}\label{BMflat}
v_{tg}(r)^{2}=\frac{r_{S}}{r}+v_{\infty }^{2}.
\end{equation}%
This tangential velocity profile could be described by the metric
\begin{eqnarray}
e^{\nu (r)} &=&\left( \frac{r}{r_{c}}\right) ^{2v_{infty}^{2}}\left( 1-\frac{%
r_{S}}{r}\right) ,  \label{generic4d2} \\
e^{\lambda (r)} &=&\frac{1}{1-\frac{r_{S}}{r}+B_{r}(r)},
\end{eqnarray}%
where $r_{c}$ is a characteristic radius of an individual galaxy. Using the
traceless property of dark radiation we obtain the differential equation for
$B_{r}$,
\begin{equation}
\left( v_{\infty }^{2}r+2r-\frac{7}{2}r_{S}\right) B_{r}^{\prime }+2\left(
1+v_{\infty }^{2}-\frac{2r_{S}}{r}\right) B_{r}+2v_{\infty }^{2}=0,
\end{equation}%
where we have used relevant approximations valid for the galactic motion. $%
B_{r}$ can be obtained by power series.

We also investigate the acceleration of slowly moving massive particle by
estimating the Christoffel symbol,
\begin{equation}
\Gamma^{r}_{tt}=\frac{1}{2r}\left(\frac{r}{r_c}\right)^{2v_{\infty}^2}\left[%
1-\frac{r_b}{r}+B_r\right]\left[2v_{\infty}^2\left(1-\frac{r_b}{r}\right)+%
\frac{r_b}{r}\right]
\approx\frac{1}{2r}\left(\frac{r}{r_c}\right)^{2v_{\infty}^2}\left(2v_{%
\infty}^2+\frac{r_b}{r}\right)\approx\frac{v_{\infty}^2}{r}+\frac{r_b}{2r}.
\end{equation}
The acceleration of the particles agree with the Newtonian gravitation
acceleration produced by a central mass and a dark matter halo.

\subsection{The metric in the power law velocity profile region}\label{powerLaw}

From Eq.~(\ref{limitc}) with $\zeta \neq 0$, we know directly one metric coefficient
\begin{equation}
e^{\nu}=A e^{\frac{v_c^2 r^{2\zeta }}{\zeta }}.
\end{equation}
In the case $\zeta < 0$, we can take the large radius approximation,
\begin{equation}
e^{\nu}\approx A\left(1-\frac{v_c^2}{\eta r^{2\eta}}\right),
\end{equation}
with $\eta=|\zeta|$. This is equivalent to a Schwarzschild correction in the asymptotic region,
\begin{equation}
B_{t}(r)\rightarrow -1+\frac{r_{\rm S}}{r}+A\left(1-\frac{v_c^2}{\eta r^{2\eta}}\right).
\end{equation}
The other correction $B_t$ can be obtained in large radius  limit from Eq.~(\ref{DEBr}), which can be simplified in this  approximation as
\begin{equation}
B_r(r)\to \frac{v_c^2}{ r^{2\eta}}+\frac{C}{r}.
\end{equation}
This metric coefficient gives the radial acceleration as
\begin{equation}
\Gamma^{r}_{tt}\approx Av_c^2r^{-(2\eta+1)}.
\end{equation}
If $A=1$, the acceleration is consistent with the acceleration produced by the dark matter density profile suggested by the power law velocity profile
\begin{equation}
\rho(r)=\frac{v_c^2(1-2\eta)}{4\pi G}r^{-2\eta-2}.
\end{equation}

\subsection{Solutions with dark pressure equation of state}

For the equation of state~(\ref{eqstate}) of the Weyl fluid, the solution of
the effective Einstein equation can be written as
\begin{equation}
e^{-\lambda \left( r\right) }=1-\frac{C_{b}}{r}-\frac{GM_{U}\left( r\right)
}{r}~,  \label{m1a}
\end{equation}%
\begin{equation}
e^{\nu \left( r\right) }\approx C_{\nu }r^{2v_{tg\infty }^{2}}\exp \left[
-C_{1}r^{-1}-C_{2}\frac{2a-3}{a-3}r^{-1+\frac{3}{2a-3}}\right] ~.
\label{m2a}
\end{equation}%
where
\begin{equation}
GM_{U}(r)\approx v_{0}r+C_{1}+C_{2}r^{\frac{3}{2a-3}}-2GM~,
\end{equation}%
and $C_{b},C_{\nu },C_{1},C_{2}$ and $M$ are also arbitrary constants of
integration. The approximate equalities indicate that in this approximation
baryonic matter (which is subdominant), was dropped. Finally the constants $%
v_{0}$ and the asymptotic tangential velocity $v_{tg\infty }$ are given in
terms of the previously introduced constants as
\begin{equation}
v_{0}=\frac{B(B-3)}{a\left( 2B-3\right) +9}~,  \label{v0}
\end{equation}%
\begin{equation}
v_{tg\infty }^{2}=\frac{1}{3}\left( av_{0}-B\right) ~.  \label{vtginfty}
\end{equation}%
This solution has been presented in detail in \cite{BraneRotCurves}.

In the weak field regime we have further simplifications, as has been
discussed in \cite{BraneRotCurves}. A careful post-Newtonian counting first
gives $B\ll a-2$, then%
\begin{equation}
v_{0}\approx v_{tg\infty }^{2}\approx \frac{B}{a-3}~,
\end{equation}%
with either $a<3/2$ and $B\leq 0$ or $a>3$ and $B>0$.

The constant $C_{\nu }$ in Eq.~(\ref{m1a}) can be immediately absorbed in
the time coordinate. We introduce the following notations
\begin{eqnarray}
\alpha  &=&\frac{3}{2a-3}~,\quad \gamma =\frac{B}{a-3}~,  \notag \\
C &=&C_{2}r_{c}^{\alpha -1}~,\quad GM_{0}=\frac{C_{1}}{2}~,
\end{eqnarray}%
and two mass type constant $M_{0}$ and $M_{b}$ given by
\begin{equation}
2GM_{1}=C_{b}+2{G}\left( M_{0}-M\right) ~,  \label{massCorrect}
\end{equation}%
The metric functions with the speed of light $c$ reintroduced can be
approximated as
\begin{equation}
e^{-\lambda \left( r\right) }\approx 1-\gamma -\frac{2{G}M_{1}}{c^{2}r}%
-C\left( \frac{r}{r_{c}}\right) ^{\alpha -1}~,  \label{grr1}
\end{equation}%
\begin{equation}
e^{\nu \left( r\right) }\approx \left( \frac{r}{r_{c}}\right) ^{2\gamma
^{2}}\exp \left[ -\frac{2{G}M_{0}}{c^{2}r}-\frac{2C}{1-\alpha }\left( \frac{r%
}{r_{c}}\right) ^{\alpha -1}\right] ~.  \label{gtt1}
\end{equation}%

In general $M_{1}$ and $M_{0}$ are two independent parameters, however $M_{1}
$ does not affect the rotation curves and should be determined by other
observation, like gravitational lensing. In the rotation curve studies $%
M_{1}=M_{0}$ has been chosen. The rotation curves explained in terms of the
Weyl fluid were characterized by three universal dimensionless constants $%
\alpha ,~\gamma ,~C$ and a mass-type constant $M_{0}$. The constant $r_{c}$
was introduced in order to have $C$ dimensionless. In LSB galaxies the
visible matter can be considered to be concentrated inside a constant
density core, as shown in \cite{deBlok}. In a stellar and gas free model,
the matching of the outside brane-world solution to the inside Newtonian
region suggests that $C=-\gamma $ and below $r_{c}$ visible matter dominates.

The constants $\alpha ,~\gamma $ obey either $\alpha <0$ or $0<\alpha <1$
(the value $\alpha =0$ is excluded, as it would correspond to the unphysical
values $a\rightarrow \pm \infty $. In both cases $\gamma $ is a small
positive number $0<\gamma \ll 1$ (with the exception of $a\approx 3$,
translating to $\alpha \approx 1$, when theoretically $\gamma $ can be an
arbitrary positive number). The choice from the analysis of the Low Surface
Brightness (LSB) galaxy rotation curves \cite{BraneRotCurves} and the
fitting from LSB galaxies suggest that $\gamma \ll 1$. The values of the
parameters $\alpha $, $\gamma $, $r_{c}$ and $M_{1}$ are presented in Table~%
\ref{Table1}.
\begin{table*}[tbp]
\begin{center}
\begin{tabular}{|c|c|c|c|c|c|}
\hline
Galaxy & $M_{0}$ & $r_{c}$ & $\alpha $ & $\gamma $ & $\chi _{\min }^{2}$ \\
\hline
& $M_{\odot } $ & kpc &  &  &  \\ \hline\hline
DDO 189 & 4.05$\times $10$^{8}$ & 1.25 & 0.3 & 6.43$\times $10$^{-8}$ & 0.742
\\
NGC 2366 & 1.05$\times $10$^{9}$ & 1.47 & 0.8 & 1.12$\times $10$^{-7} $ &
2.538 \\
NGC 3274 & 4.38$\times $10$^{8}$ & 0.69 & -0.4 & 6.73$\times $10$^{-8}$ &
18.099 \\
NGC 4395 & 2.37$\times $10$^{8}$ & 0.71 & 0.9 & 3.43$\times $10$^{-7}$ &
27.98 \\
NGC 4455 & 2.26$\times $10$^{8}$ & 1.03 & 0.9 & 2.72$\times $10$^{-7}$ &
7.129 \\
NGC 5023 & 2.69$\times $10$^{8}$ & 0.74 & 0.9 & 4.53$\times $10$^{-7}$ &
10.614 \\
UGC 10310 & 1.28$\times $10$^{9}$ & 2.6 & 0.4 & 1.12$\times $10$^{-7} $ &
0.729 \\
UGC 1230 & 3.87$\times $10$^{9}$ & 3.22 & -1.7 & 1.12$\times $10$^{-7} $ &
0.539 \\
UGC 3137 & 5.32$\times $10$^{9}$ & 3.87 & -0.5 & 1.23$\times $10$^{-7} $ &
4.877 \\ \hline
\end{tabular}%
%
%
%
%
%
%
\end{center}
\par
\caption{The best fit parameters ($M_{0}$, $r_{c}$ , $\protect\alpha $, $%
\protect\gamma $) of the sample of 9 LSB galaxies given by \protect\cite%
{BraneRotCurves}}
\label{Table1}
\end{table*}

\section{Gravitational lensing in the Weyl fluid dominated region}\label{sect5}

\subsection{Gravitational lensing in brane-world model}

In the present Section we will consider the lensing properties in the dark
radiation dominated region of a brane-world galaxy, outside the galactic
baryonic matter distribution, in the various approximations discussed in
Section IV.

\begin{figure}[b]
\begin{center}
\includegraphics[width=8cm]{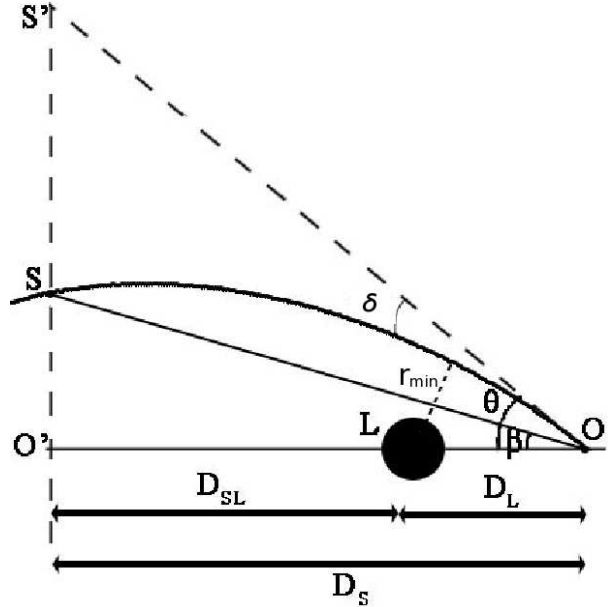}
\end{center}
\caption{The lensing geometry.}
\label{fig1}
\end{figure}

We consider the lensing geometry represented in Fig.~\ref{fig1}, with observer $O$,
lens $L$, source $S$, image $I$, and consider the deflection of light
occurring at a point $A$. Here $\theta $ is the apparent angle while $\beta $
is the real, undeflected angle of the source, and $\delta $ is the
deflection angle of the light ray. We denote by $D_{ls}$ the lens to source
distance, and by $D_{l}$ the lens distance, measured from the observer.

The deflection angle $\delta $ is obtained by comparing the two asymptotic
directions of the null geodesics \cite{WEINBERG}
\begin{eqnarray}  \label{deltadef}
\delta \left( r_{\min }\right)  &=&2\int_{r_{\min }}^{\infty }\mathcal{I}%
dr-\pi ,  \label{defl} \\
\mathcal{I}\left( r\right)  &=&\frac{1}{r}\sqrt{\frac{g_{rr}\left( r\right)
}{\left[ \left( -g_{tt}\left( r_{\min }\right) \right) /\left( -g_{tt}\left(
r\right) \right) \right] \left( r/r_{\min }\right) ^{2}-1}},  \notag \\
&&
\end{eqnarray}%
where $r_{\min }$ is the distance of minimum approach given by
\begin{equation}
\frac{dr}{d\varphi }\left( r_{\min }\right) =0.  \label{rela0}
\end{equation}

\subsection{Dark matter lensing in standard general relativity}

In the weak field limit we can use the superposition
principle, and therefore the deflection angle for light in a dark matter
model is given by
\begin{equation}
\delta =\frac{4G}{c^{2}r_{\mathrm{min}}}\left[ M+M_{\mathrm{dark}}(r_{%
\mathrm{min}})\right] ,
\end{equation}%
where $M_{\mathrm{dark}}(r)$ is the effective mass of dark matter inside
radius $r$. The existence of dark matter always enhances the deflection
angle. In the dark matter model, we can separate the contribution of the
baryonic matter and of the dark matter as
\begin{equation}
\delta =\delta _{S}+\delta _{\mathrm{dark}},
\end{equation}%
where
\begin{equation}
\delta _{\mathrm{dark}}=\frac{4GM_{\mathrm{dark}}(r_{\mathrm{min}})}{c^{2}r_{%
\mathrm{min}}}.
\end{equation}%

For example, if we consider the pseudo-isothermal halo model,
\begin{equation}
\rho _{\mathrm{iso}}(r)=\rho _{0}\left[ 1+\left( \frac{r}{r_{D}}\right) ^{2}%
\right] ^{-1},
\end{equation}%
where $\rho _{0}$ is the core halo density, and $r_{D}$ is the radius of
core halo, the asymptotic velocity of this model is a flat rotation curve, or Eq.~(\ref{BMflat}), which is obtained from the unified Schwarzschild - constant velocity metric.

\subsection{Deflection angle in the flat rotation curves region}

The deflection angle for the brane-world corrections~(\ref{asymBr}) and (\ref%
{asymBt}) can be calculated from Eq.~(\ref{deltadef}). The integration
constant $C_{t}$ does not enter the expression of $\delta $, and for the
simplest case we also set $C_{r}=0$, such that the expansion of $\delta $
for small dimensionless parameters $r_{S}/r_{\mathrm{min}}$ and $v_{\infty
}^{2}$ is given by
\begin{equation}\label{expan}
\delta =2\int_{r_{\mathrm{min}}}^{\infty }\frac{1}{r}\sqrt{\frac{r_{\mathrm{%
min}}}{r^{2}-r_{\mathrm{min}}^{2}}}\left[ 1-\right.
\left. \frac{-r^{2}+r_{\mathrm{min}}^{2}+2ln(\frac{r_{\mathrm{min}}}{r})r^{2}%
}{2(r^{2}-r_{\mathrm{min}}^{2})}v_{\infty }^{2}\right] dr-\pi .
\end{equation}%
We see from Eq.~(\ref{expan}) that the first order term in $r_{S}/r_{\mathrm{%
min}}$ vanishes, i.e., the lensing effect of the central masses are
suppressed by brane-world effects. The integral can be calculated to give
for $\delta $ the expression
\begin{equation}
\delta =\frac{3\pi v_{\infty }^{2}}{2}=\mathrm{constant}.
\label{d_constantv}
\end{equation}%
A constant deflection angle can be interpreted as resulting from a galaxy
that contains invisible mass with density profile
\begin{equation}
\rho =\frac{3v_{\infty }^{2}}{32Gr^{2}}.
\end{equation}%
The $1/r^{2}$ profile agrees with what standard dark matter models predict.
However, if we consider the deflection angle as a function of $v_{\infty }$
in the dark matter picture, a $1/r^{2}$ density halo gives
\begin{equation}
\delta _{dm}=4v_{\infty }^{2},
\end{equation}%
which is $18\%$ different from the prediction of the brane-world model. Thus
brane-world models predict a systematic deviation of the dark matter core
density deduced from lensing studies as compared to the virial mass obtained
from the study of the rotation curves.

\subsection{Deflection angle in the unified Schwarzschild - constant velocity metric}

For the brane geometry described by the metric Eq.~(\ref{generic4d2}), the
lensing deflection angle can be calculated similarly. The post Newtonian
expansion of the deflection angle is
\begin{equation}
\delta =\int \frac{1}{2r(r^{2}-r_{\mathrm{min}}^{2})}\sqrt{\frac{r_{\mathrm{%
min}}^{2}}{r^{2}-r_{\mathrm{min}}^{2}}}\left[ \left( \frac{r^{3}-r_{\mathrm{%
min}}^{3}}{r}\right) \frac{r_{S}}{r_{\mathrm{min}}}\right.
\left. -\left( r_{\mathrm{min}}^{2}-r^{2}+2r^{2}\ln \left( \frac{r_{\mathrm{%
min}}}{r}\right) \right) v_{\infty }^{2}\right] dr,
\end{equation}%
which can be numerically integrated to give
\begin{equation}
\delta =2.000\frac{r_{S}}{r_{\mathrm{min}}}+4.712v_{\infty }^{2}.
\end{equation}%
The first term is $4GM/c^{2}r_{\mathrm{min}}$, the standard GR result for
the deflection angle caused by the central mass. The second term is a
constant proportional to $v_{\infty }^{2}$, which is equivalent to
superposing on the central core an inverse square density dark matter halo.
However the proportionality constant $4.712$, which agrees with the value
given in Eq.~(\ref{d_constantv}) is different from that arising in the
standard dark matter picture, and as such it could test the model through
the correlation of the rotation curve data with lensing data.

\subsection{Deflection angle for the power law velocity profile}

For rotation curves with $v_{\rm tg}(r)=v_c r^{-\eta}$, with $\eta>0$, the metric coefficient in the braneworld model is given in Section~\ref{powerLaw}. The term of the order of $v_c^2$ in the series expansion of the deflection angle is
\begin{equation}
\delta(r_{\rm min})=v_c^2r_{\rm min}^{2\eta}\int_1^{\infty}\frac{a(1-u^2)-u^2(1-u^{2a})}{(u-1)^{3/2}(u+1)^{3/2}u^{2a+1}}du.
\end{equation}
By taking $\eta=0.3$ \cite{Sal1} we obtain
\begin{equation}
\delta(r_{\rm min})=0.345v_c^2r_{\rm min}^{0.6}.
\end{equation}
If the density and velocity profile are derived from the gravitational lensing by using general relativity, the velocity profile imply a deflection angle
\begin{equation}
\delta(r_{\rm min})=4v_c^2r_{\rm min}^{0.6}.
\end{equation}
The profile has the same form for both theories, with different characteristic constants. This is a very important consistency check of the missing mass model. If dark matter is the correct description of the missing mass problem, the velocity profiles and the lensing profiles need to be consistent.

\subsection{Lensing for models with a Weyl fluid equation of state}

In Eqs.~(\ref{grr1}) and (\ref{gtt1}), the factor $\left( \frac{r}{r_{c}}%
\right) ^{2\gamma ^{2}}$ can be approximated by 1 for $\gamma \ll 1$. As $%
M_{1}$ in general affects the lensing, we assume $M_{1}\neq M_{0}$, and we
set $M_{1}=M_{0}+M_{b}$, where $M_{b}$ is a brane-world correction to the
Schwarzschild central mass in $g_{rr}$. When $M_{b}=0$ we reobtain the
results for the rotation curves in \cite{BraneRotCurves}. Taking into account
that the exponent in Eq.~(\ref{gtt1}) is small, i.e.,
\begin{equation}
\left\vert -\frac{2{G}M_{0}}{c^{2}r}-\frac{2C}{1-\alpha }\left( \frac{r}{%
r_{c}}\right) ^{\alpha -1}\right\vert \ll 1,
\end{equation}%
and Eqs.~(\ref{grr1}) and (\ref{gtt1}) can be approximated by
\begin{equation}
e^{\nu \left( r\right) }\approx 1-\frac{2{G}M}{c^{2}r}+\frac{2\gamma }{%
1-\alpha }\left( \frac{r}{r_{c}}\right) ^{\alpha -1}~,  \label{gtt}
\end{equation}%
and
\begin{equation}
e^{-\lambda \left( r\right) }\approx 1-\frac{2{G}M}{c^{2}r}-\frac{2{G}M_{b}}{%
c^{2}r}+\gamma \left[ \left( \frac{r}{r_{c}}\right) ^{\alpha -1}-1\right] ~,
\label{grr}
\end{equation}%
respectively, where for simplicity in the following we omit the subscript of
$M_{0}$. $M_{b}$ cannot be fixed by rotation curves alone, and hence it
should be constrained by lensing observations. Notice that Eqs.~(\ref{gtt})
and (\ref{grr}) are not valid for $r\rightarrow \infty $. Since Eqs.~(\ref%
{gtt1}) and (\ref{grr1}) apply to empty space with the boundary condition
given by Eq.~(\ref{vtginfty}), the metric given by Eqs.~(\ref{gtt}) and (\ref%
{grr}) should be valid for $r_{c}<r<r_{\infty }$, where $r_{\infty}$
corresponds to the maximum spatial extension for which the geometry that
describes the rotation curves can be probed.

Expanding the integrand in Eq.~(\ref{deltadef}) to first order in both the
dimensionless parameter $\varepsilon =GM/c^{2}r_{\min }$ and the
dimensionless brane-world parameter $\varepsilon _{b}=GM_{b}/c^{2}r_{\min }$%
, and $\gamma $, we find
\begin{align}
& \mathcal{I}(r)=\mathcal{I}_{0}(r)+\varepsilon \mathcal{I}%
_{S}(r)+\varepsilon _{b}\mathcal{I}_{W1}(r)+\gamma \mathcal{I}_{W2}(r)~,
\label{deltaex} \\
\mathrm{where}  \notag \\
& \mathcal{I}_{0}(r)=\frac{1}{r}\sqrt{\frac{1}{(r/r_{\min })^{2}-1}}, \\
& \mathcal{I}_{S}(r)=\frac{r_{\min }}{r}\left( \frac{1+r/r_{\min
}+(r/r_{\min })^{2}}{1+r/r_{\min }}\right) \mathcal{I}_{0}(r), \\
& \mathcal{I}_{W1}(r)=\frac{r_{\min }}{r}\mathcal{I}_{0}(r), \\
\mathrm{and}  \notag \\
& \mathcal{I}_{W2}(r)=\frac{\mathcal{I}_{0}(r)}{2}\Bigg\{1+\left( \frac{%
r_{\min }}{r_{c}}\right) ^{\alpha -1}\times \\
& \frac{(1-\alpha )(\frac{r}{r_{\min }})^{\alpha -1}+(1+\alpha )(\frac{r}{%
r_{\min }})^{\alpha +1}-2(\frac{r}{r_{\min }})^{2}}{(1-\alpha )\left[
(r/r_{\min })^{2}-1\right] }\Bigg\},
\end{align}%
respectively. From the above series expansion it follows that the deflection
angle can be decomposed into the Schwarzschild contribution and the
brane-world contribution,
\begin{equation}
\delta =\delta _{S}+\delta _{W}.  \label{deltasw}
\end{equation}%
$\delta _{S}$ can be integrated by means of the substitution $u=r/r_{\min }$
(then also $du/u=dr/r$), so that
\begin{equation}
\delta _{S}=2\int_{r_{\min }}^{\infty }\left[\mathcal{I}_{0}(r)+\varepsilon
\mathcal{I}_{S}(r)\right]dr-\pi =\frac{4GM}{c^{2}r_{\min }},  \label{deltas}
\end{equation}%
$\delta _{W}$ can be written as
\begin{equation}  \label{deltaw}
\delta_W=2\int_{r_{\min }}^{\infty }\left[\varepsilon_b\mathcal{I}%
_{W1}(r)+\gamma\mathcal{I}_{W2}(r)\right]dr
=\frac{2GM_b}{c^2r_{\min}}+\gamma\left[\frac{\pi}{2}+\left(\frac{r_{\min}}{%
r_c}\right)^{\alpha-1}g(\alpha)\right],
\end{equation}
where
\begin{equation}
g(\alpha)=\int_{1}^{\infty }\frac{du}{u}\sqrt{\frac{1}{u^{2}-1}}
\left[\frac{(1-\alpha )u^{\alpha -1}+(1+\alpha )u^{\alpha +1}-2u^{2}}{%
(1-\alpha )\left( u^{2}-1\right) }\right],
\end{equation}
is a function depending on $\alpha$ only.

\subsection{Lensing deflection angle}

\label{criticalrsection}
The contribution of the brane Weyl curvature in Eqs.~(\ref{gtt}) and (\ref%
{grr}) (the term proportional to $\gamma $) makes $\left( -g_{tt}\right) $
to increase, while it makes $g_{rr}$ to decrease. Therefore the contribution
of the terms containing $\gamma $ to $\mathcal{I}$ can be of either sign.
This result can be verified by using the series expansion in Eq.~(\ref%
{deltaw}). The positive or negative sign of $\delta _{W}$ depends on the
parameters of the system. The effect of dark matter to the lensing is
mimicked by the brane-world effects when $\delta _{W}$ is positive. Fig.~\ref%
{deltaW_rmin_diff_alpha} shows how $\delta _{W}$ can be positive or negative
in the parameter space.

\begin{figure}[b]
\begin{center}
\includegraphics[width=8cm]{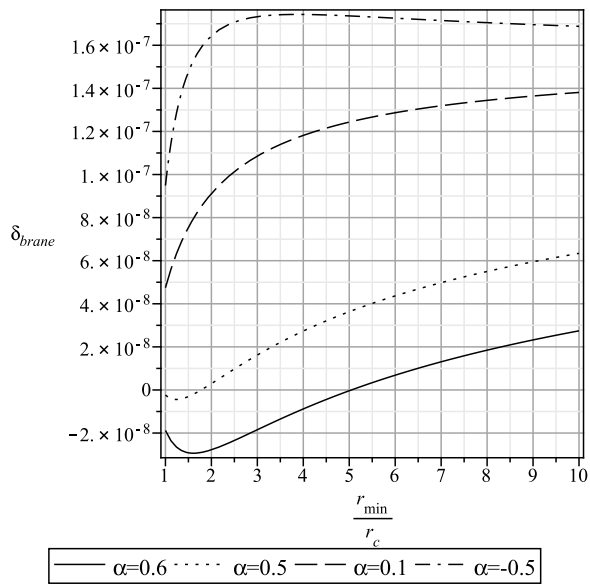}
\end{center}
\caption{The plot shows the evolution of the braneworld contribution $%
\protect\delta _{W}$ to the deflection angle, as function of $r_{\min
}/r_{c} $ for various values of the braneworld parameter $\protect\alpha $.
The contribution $\protect\delta _{W}$ is positive for orbits with the
closest approach $r_{\min }$ above the critical radius $r_{\min }^{crit}$
(defined by the intersection with the horizontal axis).}
\label{deltaW_rmin_diff_alpha}
\end{figure}

The existence of a negative $\delta _{W}$ in the parameter space is a unique
feature of the brane-world models as compared to the conventional dark
matter models. Standard dark matter models assume an invisible mass
distribution surrounding the galaxy, with the dark matter particles feeling
the gravitational interaction, but interacting weakly with baryonic matter
in other interactions. The existence of dark matter should always enhance
the gravity, and thus increase the deflection angle relative to a "visible
matter" lens.

\section{Gravitational lensing and rotation curves for LSB galaxies}\label{sect6}

The braneworld Weyl fluid model defined by Eq.~(\ref{eqstate}) can  be
discriminated with respect to standard dark matter models by
lensing studies. In \cite{BraneRotCurves}, the rotation curves of the LSB galaxies were fitted with the Weyl fluid model. We would like to compare standard dark matter lensing with braneworld lensing by using the data obtained from the corresponding rotation curves fitting in dark matter and braneworld models.

\subsection{The rotation curves fitting}

The rotation curves of the LSB galaxies used in \cite{BraneRotCurves} have been obtained in \cite{deBlok}. As an example of the standard dark matter approach we consider the pseudo-isothermal halo model discussed in \cite{deBlok}. The values of the parameters $\rho _0$ and $r_D$ obtained with the
minimum-disk assumption for our sample of 9 LSB galaxies are presented in \cite{deBlok}. However, for the comparison of the observational data with the brane world model predictions we only need the qualitative result that dark matter always gives positive deflection angle, i.e. it increases the deflection caused by the central mass.

\begin{figure}[h]
\includegraphics[width=9cm]{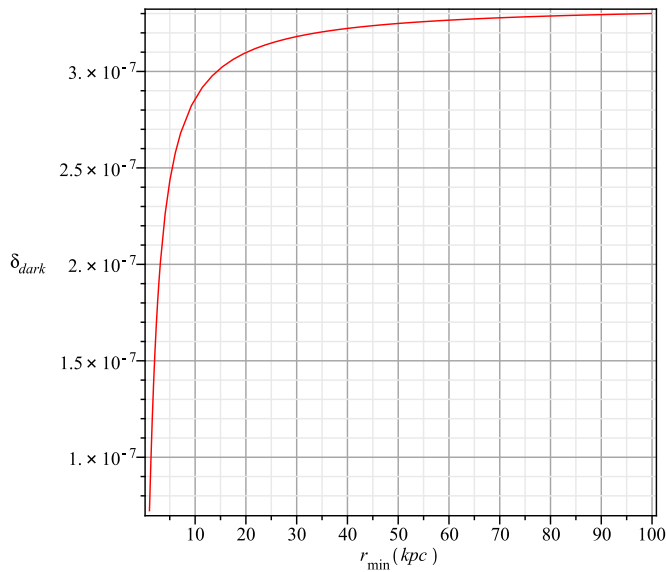}
\caption{The standard dark matter model deflection angle $\protect\delta_{%
\mathrm{dark}}$ at different impacts $r_{\mathrm{min}}$ predicted by the
pseudo-isothermal halo model, with $\protect\rho_0=10^{-17}\mathrm{g\;cm^{-3}%
}$ and $r_c=1\;\mathrm{kpc}$. We only need the result that the deflection angle is always positive.}
\label{deltaDM}
\end{figure}

Fig.~\ref{deltaDM} represent the lensing profile of the chosen sample of
galaxies. The shape of the plots are similar for all galaxies in the sample,
and we pick $\rho_0=10^{-17}\;\mathrm{g\;cm^{-3}}$ and $r_c=1\;\mathrm{kpc}$
for illustration.

\subsection{Brane-world deflection}

We first define the critical radius of approach $r_{\min }^{crit}$ by $%
\delta _{W}(r_{\min })=0$. The lensing by Weyl fluid could be sufficiently
different from that produced by dark matter if lensing occurs inside $%
r_{\min }^{crit}$, as can be seen from comparing Fig.~\ref%
{deltaW_rmin_diff_alpha} with Fig.~\ref{deltaDM}.

The value of $r_{\min }^{crit}$ is given by the equation
\begin{equation}
\frac{2r_{b}}{r_{\min }^{crit}}+\gamma \left[ \frac{\pi }{2}+\left( \frac{%
r_{\min }^{crit}}{r_{c}}\right) ^{\alpha -1}g(\alpha )\right] =0.
\label{critical}
\end{equation}%
The location of $r_{\min }^{crit}$ depends on $r_{b}$, $\alpha $, $r_{c}$
and $\gamma $. From these parameters, only $r_{b}$ cannot be fixed by the
rotation curve analysis. If $r_{\min }^{crit}$ is located inside the range
of validity of the metrics given by Eq.~(\ref{gtt}) and Eq.~(\ref{grr}),
there can be hope for detecting it.

In order to investigate the possibility of observing lensing with $r_{\min
}<r_{\min }^{crit}$, we pick galaxies DDO189, NGC3274, NGC2366 and NGC4455
for analysis. There are two possibilities that could arise from the sample
presented in Table \ref{Table1}. We use the Weyl fluid parameters fixed by
rotation curve studies \cite{BraneRotCurves}, and plot $\delta _{W}$ versus $%
r_{\mathrm{min}}$ for different $r_{b}$ (in units of $10^{-8}r_{c}$). The
solid line represents the case $r_{b}=0$. There are two cases, illustrated
on Figs. 4 and 5.

\paragraph{$r_{\min }^{crit}\lessapprox r_c$}

In this case $r_{\min }^{crit}$ is well hidden inside the galaxy, and
therefore gravity is not weakened, but rather enhanced anywhere in the
asymptotic region. Brane-world contributions and dark matter give similar
predictions.
\begin{figure}[h]
\includegraphics[width=9cm]{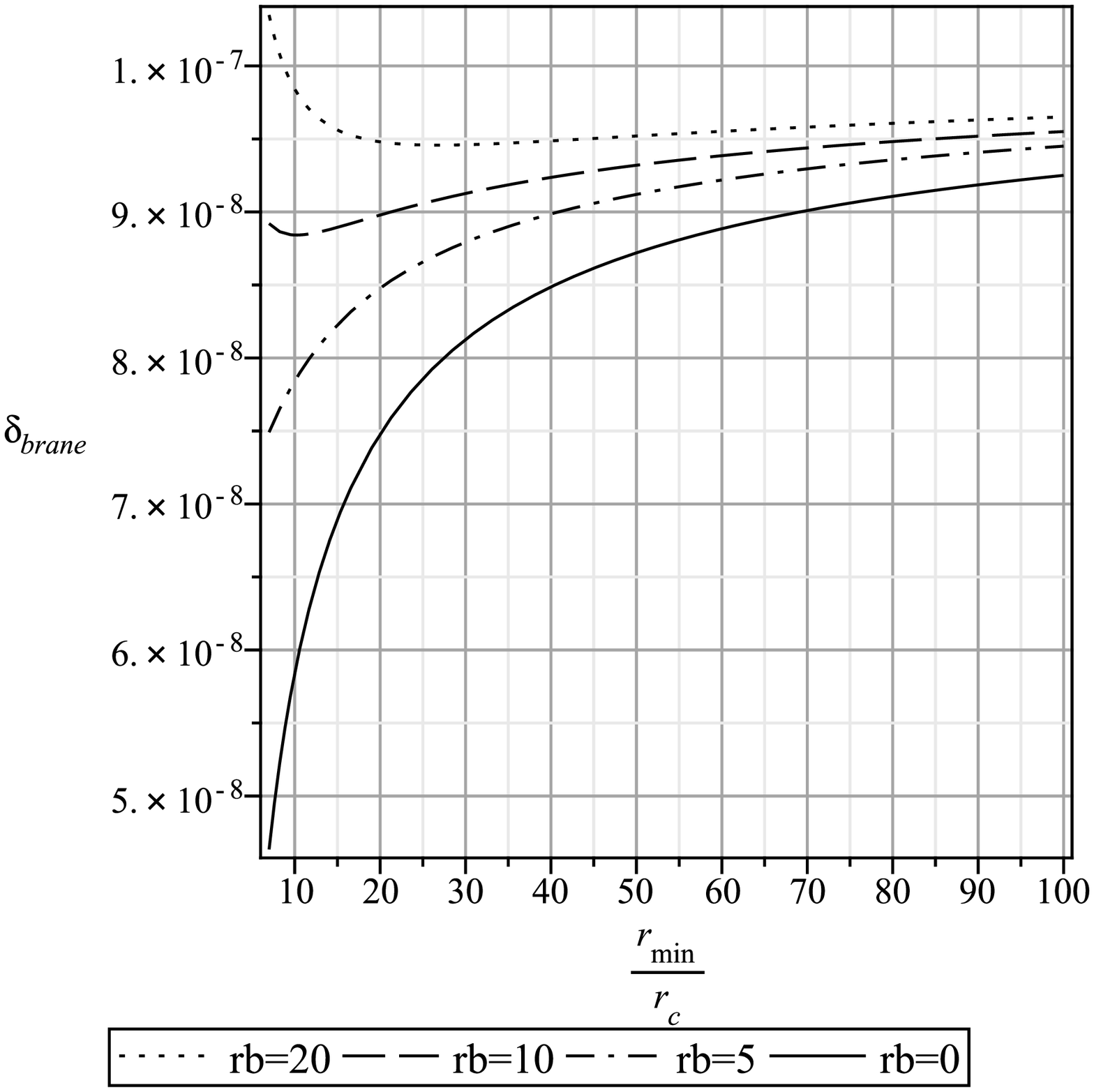}
\caption{The lensing profile of DDO 189 with different $r_{b}$ (in unit of $%
10^{-8}r_{c}$). Rotation curve data extended to 9kpc, which is similar to 7$%
r_{c}$.}
\label{ddo189}
\end{figure}
\begin{figure}[h]
\includegraphics[width=9cm]{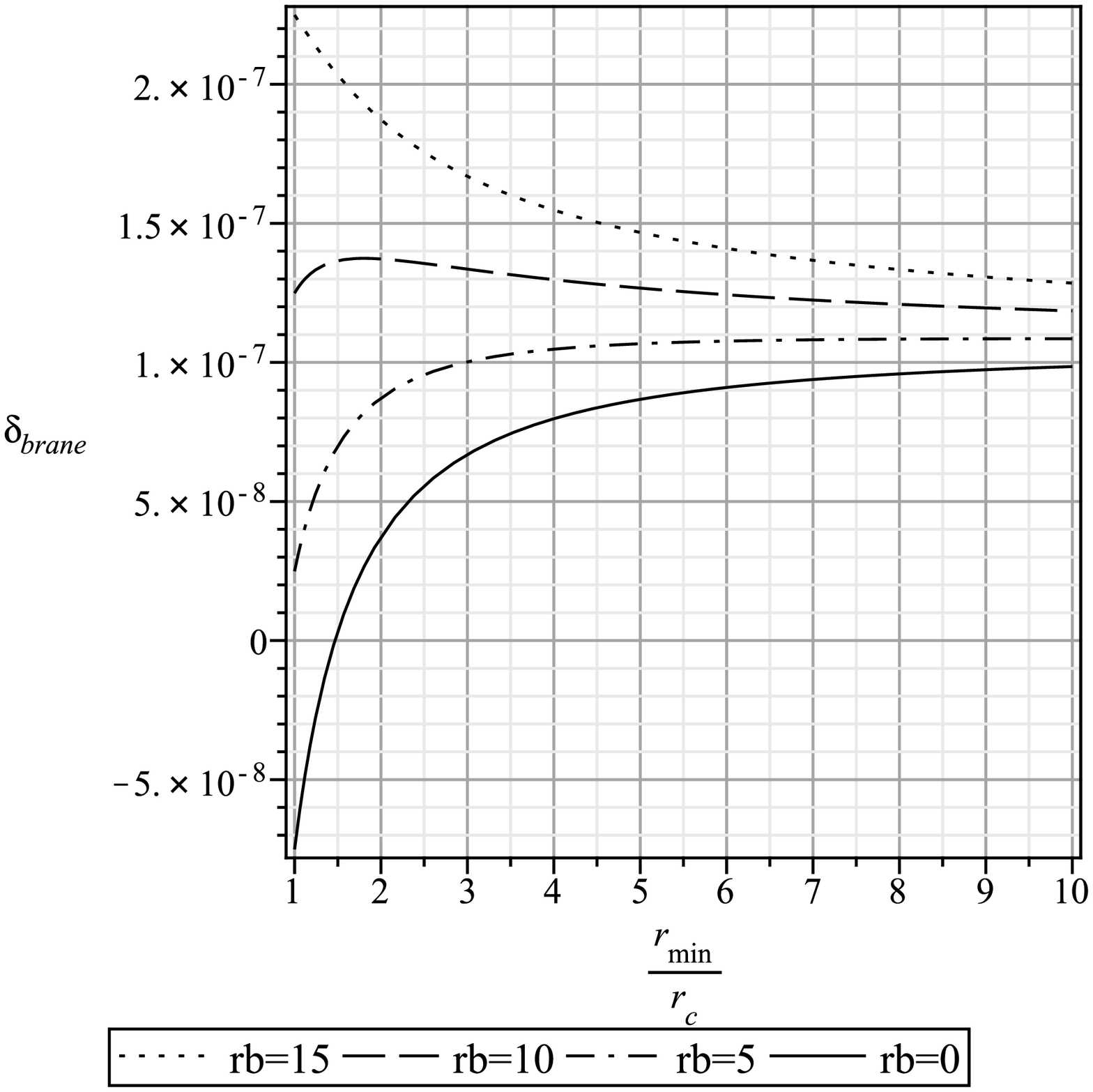}
\caption{The lensing profile of NGC 3274 with different $r_{b}$ (in unit of $%
10^{-8}r_{c}$). Rotation curve data extended to 9kpc, which is similar to 7$%
r_{c}$.}
\label{ngc3274}
\end{figure}

\paragraph{$r_{\min }^{crit}> r_c$}

In this case $r_{\min }^{crit}$ could be detected by astronomical
observations. The lack of observations of such lensing effects may possibly
suggests that the choice of the equation of state for the Weyl fluid given
by Eq.~(\ref{eqstate}) may not be appropriate for the respective galaxies.

\begin{figure}[h]
\includegraphics[width=9cm]{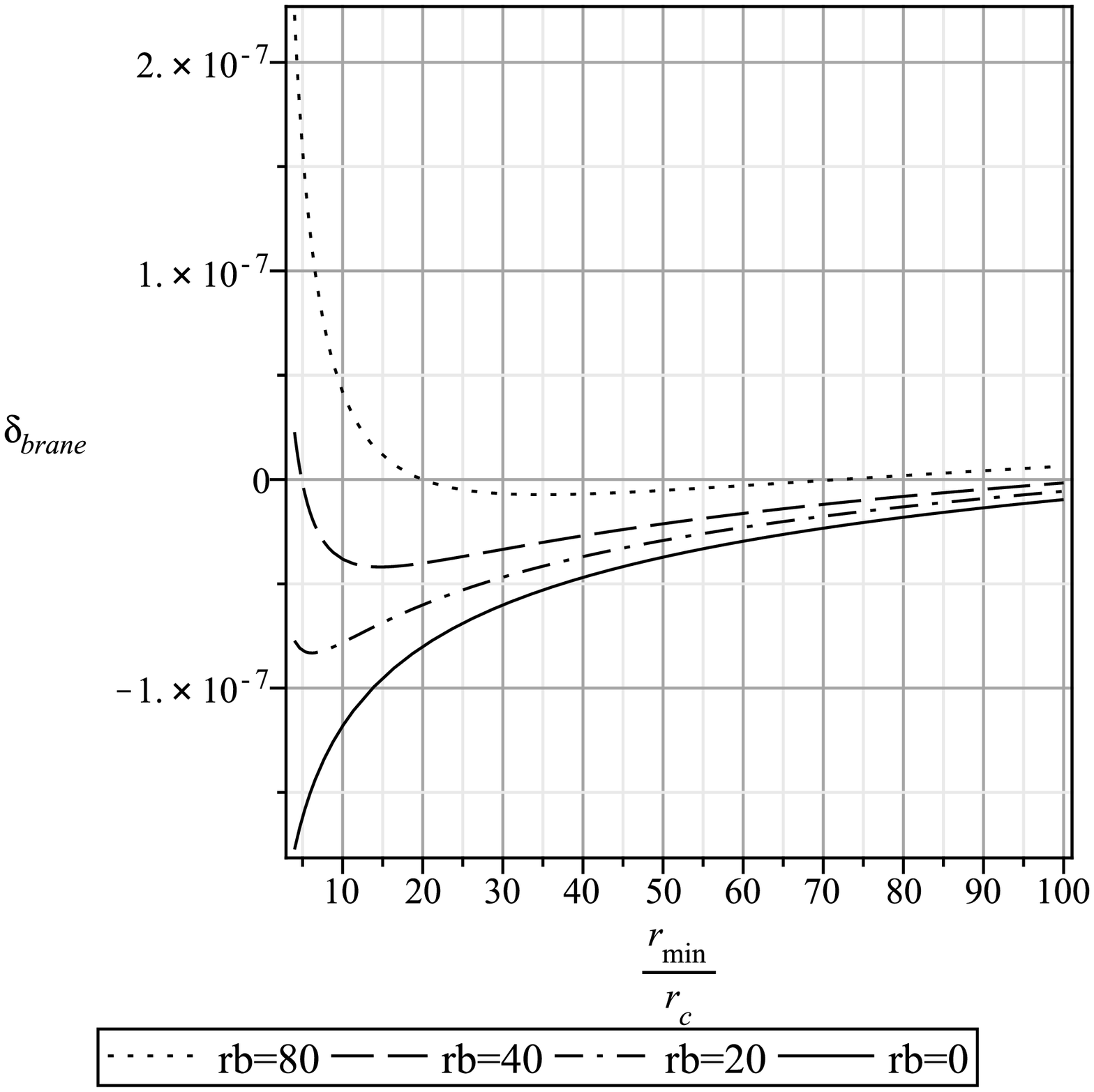}
\caption{The lensing profile of NGC 2366 with different $r_{\rm S}$ (in unit of $%
10^{-8}r_c$). Rotation curve data extended to 6kpc, which is similar to 4$%
r_c $.}
\label{NGC2366}
\end{figure}

\begin{figure}[t]
\includegraphics[width=9cm]{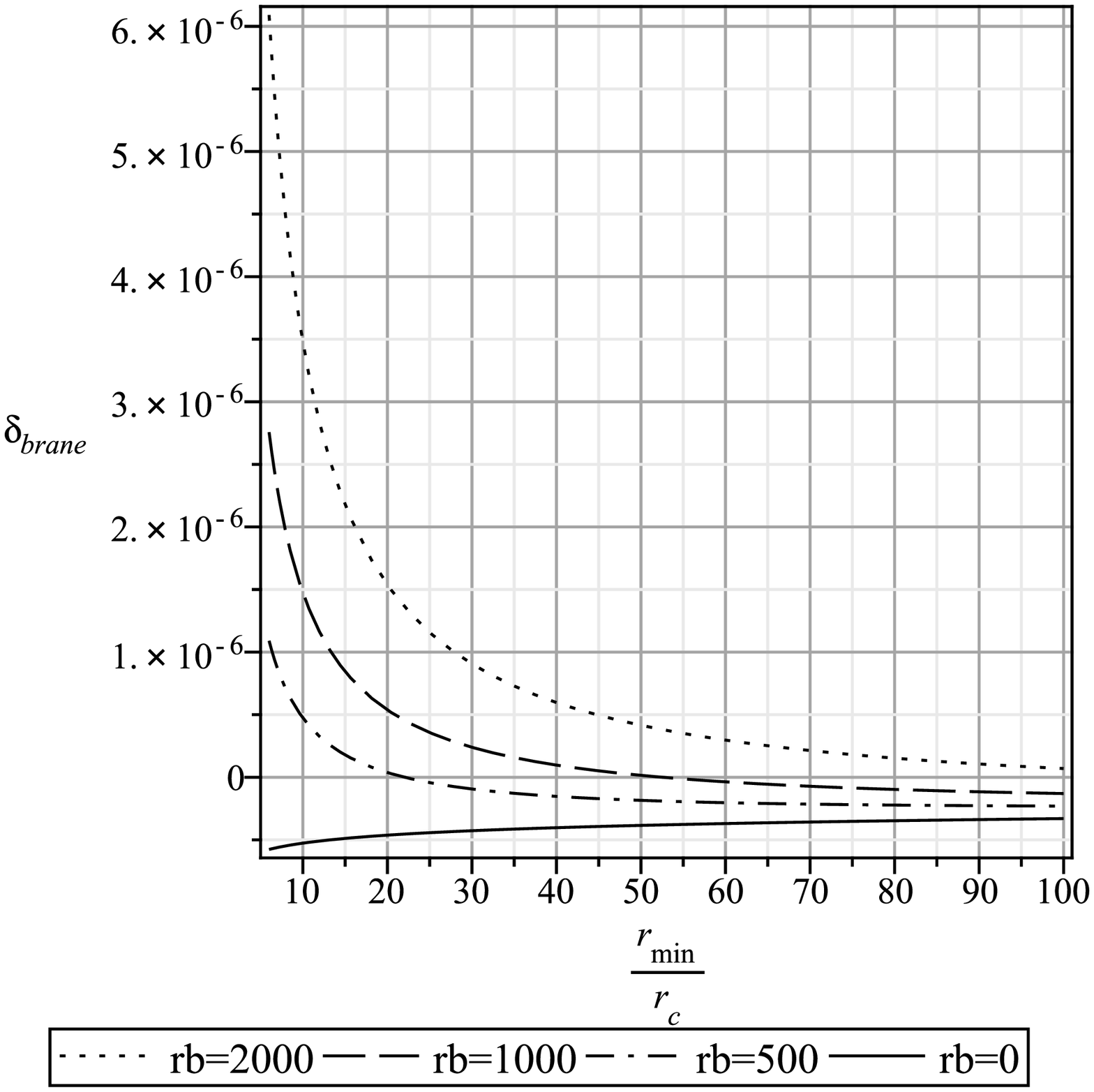}
\caption{The lensing profile of NGC 2366 with different $r_{\rm S}$ (in unit of $%
10^{-8}r_c$). Rotation curve data extended to 6 kpc, which is similar to 4$%
r_c $.}
\label{NGC4455}
\end{figure}

\section{Discussions and final remarks}\label{sect7}

Dark matter searches have been unsuccessful so far. Therefore attempts to
explain galactic rotation curves and galactic cluster dynamics by modifying
general relativity at large scale cannot be excluded \textit{a priori}. The
spherically symmetric brane-world model offers a solution for the missing
mass problem without the need to introduce unidentified forms of matter.
Whether such a model is viable, can be decided by working out a number of
its predictions and confronting the theoretical predictions with
observations. A consistency check on rotation curves and lensing data,
proposed in the present paper, is therefore relevant for this purpose.

This paper has shown that specifying an induced metric on the brane yields a
unique solution of the bulk in this particular problem. This happens because
we only consider modification of gravity in the dark radiation dominant
region, i.e. the region in which induced metric is analytic. The schemes
described in the Campbell-Magaard and Cauchy-Kowalewski theorems allow to
locally embed the brane into a region of 5d space-time with negative
cosmological constant. Therefore we could employ the rotation curve data of
galaxies to fix the brane metric. However an asymptotic AdS5 bulk geometry
is unlikely.

The corrected metric arising in the brane-world model could explain
observations related to the motion of massive particles in stable circular
orbits around galaxies. Traceless dark radiation could indeed behave like
dark matter. The model could in principle also explain the missing mass in
gravitational lensing without assuming dark matter.

However, by correlating the rotation curve data and gravitational lensing,
the brane-world models show a distinctive feature over standard dark matter
models. The proportionality constant between the limiting constant
tangential velocity square and the deflection angle are biased by $18\%$.
This effect could be however blurred by the error margins in the rotation
curve data. The bias is valid for all galaxies with a flat rotation curve.
Although the effect is hard to detect in individual galaxies, one can still
do a survey on rotation curve-lensing correlation, and investigate the best
fitting result.

The unified Schwarzschild - constant velocity metric represents a different
way of fixing the brane-world metric by using the properties of the rotation
curves. It presents another expression for the tangential velocity limit of
the rotation curves. The present study suggests that one could use simple
analytical functions to fit the rotation curves. It consolidates the
proposal of using observational data on the rotation curves as a probe of
the existence of the extra dimensions. The predictions of the lensing in the
presence of extra-dimensional effects are consistent with the limiting
constant tangential velocity calculations. Particle motion can be described
by introducing an effective mass component, but there is some disagreement
in the rotation curve and lensing correlation.

In the model with dark radiation equation of state, we have found that there
are certain parameters of the model that mimic the observed lensing profiles
of the dark matter halos. Galaxies DDO189 and NGC3274 give enhanced gravity
in all cases. Some numerical values of the parameters in this model however
could produce a negative contribution to the deflection of light, like, for
example, in the case of the galaxies NGC2366 and NGC4455. The negative
contribution to deflection is due to the repulsive effect of the brane-world
gravity; and we consider this an indication that the dark radiation equation
of state may be inappropriate. The fact that we always found gravitational
enhancement in galaxies may rule out the choice of the linear equation of
state of the Weyl fluid for these galaxies.

On the other hand, we found that a traceless energy momentum tensor on the
brane could explain astronomical observations independently of any equation
of state of the dark radiation, which could solve the missing mass problem.
Such an effective fluid could have pressure, but does not have any free
streaming. Conventional constraints on dark matter that assume that dark
matter has to be cold due to the free streaming processes in structure
formation history may not apply to the case of the effective Weyl fluid.

Obtaining the full extra dimensional metric is among the biggest challenges
in brane-world models. Further constraining the equation of state for the
dark radiation could also help to better understand both the extra
dimensional features of brane-worlds, and also the properties of dark matter.

\acknowledgments

KCW wishes to acknowledge valuable discussions with Shinji Mukohyama and
Masamune Oguri on brane-worlds and lensing. L\'{A}G is grateful to Tiberiu
Harko and Kwong Sang Cheng for hospitality during his visit at the
University of Hong Kong, during which this work has been initiated and to
Sanjeev Seahra for discussions. The work of TH was supported by RGC grant
No.~7027/06P of the government of the Hong Kong SAR.

\end{document}